Considerations for health care institutions training large language models on electronic health records

Weipeng Zhou; Danielle Bitterman, MD; Majid Afshar, MD, MSCR; Timothy A. Miller, PhD

**Abstract**

Large language models (LLMs) like ChatGPT have excited scientists across fields; in medicine, one source of excitement is the potential applications of LLMs trained on electronic health record (EHR) data. But there are tough questions we must first answer if health care institutions are interested in having LLMs trained on their own data; should they train an LLM from scratch or fine-tune it from an open-source model? For healthcare institutions with a predefined budget, what are the biggest LLMs they can afford? In this study, we take steps towards answering these questions with an analysis on dataset sizes, model sizes, and costs for LLM training using EHR data. This analysis provides a framework for thinking about these questions in terms of data scale, compute scale, and training budgets.

1. Introduction

The recent release of ChatGPT by OpenAI [1], along with subsequent releases of large language models (LLMs) by OpenAI (GPT-4) [2], Anthropic [3], [4], Google[5], Meta AI [6], [7], and other groups [8]–[10], has led to incredible enthusiasm for generative large language models. These models typically learn from vast corpora of text data, have billions of tunable parameters, and also often make use of smaller amounts of human-curated data for suitability for downstream tasks. At their best, they can answer questions on diverse topics with highly fluent and informative responses, solve problems they are given in the input prompt, display seemingly coherent chains of reasoning, and act as conversational search engines.

While LLMs have attained great popularity of late, creating LLMs requires resources and algorithms that are not widely understood (see Figure 1 for a diagram of the training process).

"Pre-training" is the most intensive part of the learning algorithm, and involves prompting a large model (typically a transformer with a decoder-only architecture) to predict the next word that will occur in a sequence, and using back-propagation to adjust model weights to improve at the task. This process uses datasets of general domain data such as the Colossal Clean Crawled Corpus (300 GB) [11], The Pile (800 GB) [12] and RefinedWeb (2.8 TB) [13]. Over datasets such as these with billions or trillions of words of text, the model learns internal representations that form the basis of an LLM. This work is primarily concerned with that pre-training step, but aspects of the other steps may need to be considered in light of our findings. Table 1 introduces some definitions to ground the discussion in this work.

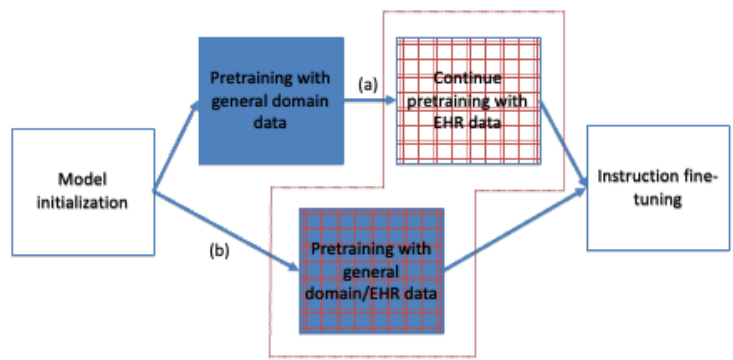

*Figure 1: An overview of possible training approaches for incorporating electronic health record (EHR) data that this work discusses.*

Better understanding of the training of LLMs is important because LLMs could have a large potential impact in the field of healthcare, as other recent work has acknowledged [14]–[16]. Promising recent applications include summarizing medical evidence [17] and a chatbot for answering psychiatry patient questions [18]. But more generally, in the same way that LLMs can ingest data about celebrities from the world wide web during pre-training and then answer questions about them, perhaps they could also ingest entire EHRs – training on the dozens or hundreds of notes that are available for many patients, across populations of millions of such patients, to summarize, answer questions, or retrieve certain data points about specific patients. Just as they seem to display reasoning ability in mathematics and word problems by encountering many such explanations in their pre-training data, they may be able to infer clinical reasoning based on decision-making documented in EHRs to provide diagnostic and treatment decision support.

*Table 1. Important definitions for terms related to large language models used in this work.*

| NLP concept | Definition |
|---|---|

| Transformer | Transformer is the state-of-the-art neural network architecture for natural language processing. It consists of an encoder and/or a decoder. Before feeding a word sequence into the model, a tokenizer will be used for preprocessing. |
|---|---|
| Generative model | In the current context, these are typically instances of a pre-trained decoder-only transformer used for generating text. It receives a free text "prompt" as input and generates next word sequences auto-regressively. |
| Large language model (LLM) | Typically refers to generative models with billions of parameters that are trained on large text corpora. LLMs are general purpose and can handle various tasks such as question answering, inference, and summarization. |
| Pretraining | Pretraining is the initial phase of training a model where the model is exposed to a large dataset. In generative LLMs, the pretraining task is to predict the next word in a sequence given the context provided by the previous words. The model updates its parameters based on the prediction error. |
| Continued pretraining | Continued pretraining refers to the process of further training a pretrained LLM on an additional, domain specific dataset (e.g., EHR). It allows the model to continue learning and adapt to domain specific knowledge. |
| Instruction fine-tuning | Instruction fine-tuning involves updating the parameters of an LLM using a smaller amount of data that is aligned with specific human needs. Training data for instruction fine-tuning can include human produced input-output pairs or human rankings of sampled model responses. |

Tantalizing use cases such as these have led to unprecedented enthusiasm for natural language processing (NLP), the field of computer science concerned with programming computers to understand human language, across the health care world. However, the logistics of creating and deploying such models are beyond the understanding even of many NLP researchers. There are important findings from the LLM literature around training data size that have not penetrated widely into the clinical informatics world. Specifically, recent work has derived "scaling laws" [19] that approximate the size of model and amount of data required for pre-training LLMs.

There are also important unresolved issues that will apply to EHR data. One important aspect of EHR data compared to web data is that it will not have as many explicit examples of people summarizing things, asking and answering questions, or providing explicit reasoning.

While instruction fine-tuning [20] has shown a remarkable ability to train LLMs to produce outputs that are aligned with human expectations with a relatively small amount of supervised data, that ability is likely leveraging similar examples from the pre-training data. This leads to unanswered questions including: Is there value in institutions training their own LLMs (mostly) from scratch? What are the tradeoffs between training from scratch and doing continued pre-training from an already trained base model? Here, we take steps towards answering these questions with an analysis on dataset sizes and costs for LLM training using EHR data. This analysis provides a framework for thinking about these questions in terms of data scale, compute scale, and costs.

2. **Objective**

The objective of this manuscript is to outline known and unknown issues around the training of LLMs, and specifically around data and compute requirements, to help decision-makers at healthcare institutions better understand opportunities and challenges for LLM development and deployment.

3. **Methods**

*3.1 Approximations and assumptions*

The transformer architecture consists of multiple layers of identical blocks. By scaling the number of these blocks, the dimensionality of the components inside the blocks, and even the number of word types in the vocabulary, among other factors, the number of learnable parameters in a transformer-based model can be easily scaled. In general, larger parameter counts lead to better downstream performance – for example, the general-domain leaderboards are dominated by larger models [21], and in the medical domain larger models perform better on USMLE tests [22]. However, the more parameters one wants to use in a model, the more data is required to train it. Hoffman et al. [19] experimented with model and data scale in the training of LLMs, and estimated that, to optimize training loss, the model size and the number of training "tokens" should scale together (i.e., to double the number of model parameters, one should double the size of the input data as well).[1] They then showed that these empirical estimates could be used to guide the optimal training of an LLM, named "Chinchilla," that had many fewer parameters than a previous model, "Gopher," but obtained similar performance. The LLaMA [6] models attempted to follow these Chinchilla scaling laws, training models as large as 65 billion parameters (hereafter *LLaMA-65b*) using 1.4 trillion tokens (the recently released LLaMA-2 models [7] were trained on more tokens, representing a strategy of producing optimized models of certain sizes rather than optimizing a fixed compute amount, so we focus our analysis on the first LLaMA releases). The compute requirements for a single training run were 2048 A100 80G GPUs for approximately 21 days, resulting in 1,022,362 GPU hours. Since the LLaMA technical report described extensive training statistics in terms of Nvidia A100 80G, which is the GPU model widely used in the public and private sectors at the time of paper writing [6], [7], [16], [24], the estimates in this manuscript use statistics based on that GPU as a foundation.

*3.2.1 Model size, token size, and computational cost*

Using the Chinchilla findings [19], one can obtain a relationship between the model size and the optimal training token size, represented by the equation:

$$\log_{10}(\text{model size}) = 0.9606 * \log_{10}(\text{number of tokens}) - 0.8981 \quad (\text{Equation 1})$$

Computer scientists use FLOPs (floating-point operations) as a unit to quantify the number of computations that are performed by a specific algorithm. Kaplan et al. [25] found that the computational cost for training LLMs can be estimated as:

$$\text{cost}_{\text{FLOPs}} = 6 * \text{model size} * \text{token size} \quad (\text{Equation 2})$$

---

[1] Because LLMs have limited vocabularies, rare words may be broken down into multiple "word pieces" and the input sequence length is measured in terms of the number of these word piece tokens. In English, a token is approximately equivalent to 0.75 words[23].

*3.2.2 Computational cost, computing time, training budget*

One can also establish a relationship between computational cost (in FLOPs) and computing time (in hours), based on Touvron et al. [6]. As noted above, the LLaMA-65b model was trained with 1.4 trillion tokens using 1,022,362 GPU hours. Applying equation 2, we find that the computational cost for LLaMA-65b is 6 * 65 billion * 1.4 trillion FLOPs. Dividing this by the reported 1,022,362 GPU training hours, we can estimate the processing power of an A100 80G GPU for LLM training as 5.35701E+17 FLOPs per hour. Cloud GPU hourly rates can vary widely from different providers, instance types, and tend to decline over time, but in this work we use the current rate for renting an A100 80G GPU on Amazon Web Services (AWS) of approximately $3 per hour (1-year reserved rate, $5 per hour if on-demand) [26]. Therefore, we can estimate:

computing time = $\text{cost}_{\text{FLOPs}}$ / GPU processing power (Equation 3)

training budget = computing time * GPU price (Equation 4)

*3.2.3 Token size and dataset size*

LLMs read their inputs as sequences of word "tokens," which are word primitives that can be combined to represent nearly all potential inputs. For example, a complex word like "colonoscopy" will be modeled to an LLM as a sequence of sub-word tokens "colon", "os" and "copy". These tokens are the units that scaling laws use to measure dataset size. Most institutions will not know how many training "tokens" they have, but may be able to easily check how large their text data is on disk, so we frame our methods based on this latter measurement. To estimate the relationship between these two quantities, we used the same tokenizer used for LLaMA, and tokenized medical notes from University of Washington Medical Center (UWMC), Boston Children's Hospital (BCH), and MIMIC-III [27] (see Table 2 for dataset details). We ran the tokenizer to get a token count for each dataset, then divided the number of tokens by the size on disk to get an estimate of EHR-derived

tokens/gigabyte on disk. We can then convert the training token size to gigabytes using the Equation:

$$\text{dataset size} = \text{token size} / \text{tokens/gigabyte} \quad \text{(Equation 5)}$$

We use the estimation methods described above to create tables from two perspectives. The first perspective is of an institution that wants to pre-train a model from scratch using either EHR data only, or combination of general domain and EHR data. Given a well-known target model that one might wish to replicate with EHR data (e.g., LLaMA-65B, ChatGPT), we estimate how much data is required, and how much training would cost to allow for one training epoch (the exact mix of general-domain and EHR data to use is an unresolved question). The second perspective is of an institution that may want to take an existing pre-trained model (e.g., LLaMA-65B) and pre-train it further using their EHR of a certain size.

## 4. Results

First, we find that that in the MIMIC-III, UW and BCH datasets, the token-to-GB ratios are 0.40, 0.35, and 0.34 billion tokens per gigabyte (Table 2). We used the middle value, 0.35, in Equation 5 in computing subsequent results. We use the University of Washington Medical Center as a reference throughout; with 37,000 annual hospital admissions, and almost 1 million outpatient and emergency visits per year, initial survey response from the UW Clinical Research Data Warehouse estimated 60 GB per year in EHR text data generated per calendar year.

*Table 2. Description for the MIMIC-III, UW and BCH dataset.*

| Dataset name | Data source | Disk size | Setting | Token size (Billion) | Token size/Disk size (Billion/GB) |
|---|---|---|---|---|---|
| MIMIC-III | All note types from patients admitted to the critical care units of Beth Israel Deaconess Medical Center. | 3.8GB | Inpatient | 1.53B | 0.4 |
| UWMC | All note types from out of hospital cardiac arrest (OHCA) patients, as well as patients matched by age and sex across the University of Washington healthcare system | 4.1GB | Inpatient and outpatient | 1.45B | 0.35 |

| | | | | | | |
|---|---|---|---|---|---|---|
| BCH | All note types from patients in an adult congenital heart disease cohort | 1.6GB | Inpatient and outpatient | 0.55B | 0.34 | |

Table 3 presents the required token size, corresponding dataset size, and budget cost for one training epoch, based on several reference model sizes. We note that for a model the size of GPT-2 (1.5 billion parameters) [28], the dataset size requirement is 87.7 GB, and the cost of one training epoch on the data is around $1,500. For a GPT-3/ChatGPT-size model (175 billion parameters) [1], [29], the optimal dataset size requirement is 12.4 TB, costing more than $25 million for one pass through the data. Table 3 represents models trained from scratch, so that the required token counts represent combined counts of EHR and other sources of data.

*Table 3.* Model parameter size, example model, optimal training token size, corresponding database size, cost and training time for one training run on a single rented AWS NVIDIA A100-80G GPU. Assuming 365 days per year.

| Model size (Parameters) | Example model | Optimal training token size | Database size | Cost (USD) | Training time |
|---|---|---|---|---|---|
| 1.5B | GPT-2 | 30.7B | 87.7GB | $1.5K | 21.5 days |
| 7B | LLaMA-7B | 152.6B | 436.1GB | $35.9K | 1.4 years |
| 13B | LLaMA-13B | 290.8B | 830.8GB | $127.0K | 4.8 years |
| 33B | LLaMA-33B | 766.8B | 2.2TB | $850.3K | 32.4 years |
| 65B | LLaMA-65B | 1.6T | 4.4TB | $3.4M | 129.1 years |
| 175B | GPT-3/ChatGPT | 4.4T | 12.4TB | $25.6M | 974.3 years |

Table 4 shows the budget of pre-training models of various sizes using a varying amount of data, representing the cost of the alternative technique of applying continued pre-training to a large model that has already been pre-trained. The 65 billion parameters corresponds to some of the largest available open-source models available at the time of writing this manuscript [6].

If we assume UWMC has been generating data at the same rate for 10 years, that corresponds to 600GB. We can look in the table at the closest value in the Table 4 (model size=65B, database size=500GB), and see that the cost to get through one pass of the data for a 65 billion parameter model is around $382.2K. At the bottom of the table, we see that to continue pre-training with 10 TB, which might be possible at a massive institution or a consortium of institutions, the estimated cost to get through the data one time is $7.6 million.

*Table 4*. *The cost of continuedly pretraining models on databases with different sizes.*

| | Cost (USD) | | | | |
|---|---|---|---|---|---|
| Model size* (→) | 1.5B | 7B | 13B | 33B | 65B |
| Database size (↓) | | | | | |
| 20GB | $352.8 | $1.6K | $3.1K | $7.8K | $15.2K |
| 100GB | $1.8K | $8.2K | $15.3K | $38.8K | $76.4K |
| 500GB | $8.8K | $41.2K | $76.4K | $194.0K | $382.2K |
| 1TB | $17.6K | $82.3K | $152.9K | $388.1K | $764.4K |
| 5TB | $88.2K | $411.6K | $764.4K | $1.9M | $3.8M |
| 10TB | $176.4K | $823.2K | $1.5M | $3.9M | $7.6M |

*Model size refers to the size of the model initiated for continued pre-training.

## 5. Discussion

Our results can be used to make sense of a few different scenarios we anticipate:

*Evaluating vendor offerings*

Technology vendors with cloud and artificial intelligence (AI) offerings have already made LLM-based products available to health care institutions, and their offerings will likely soon include customized training for large language models [30]. Evaluating the upside potential of these tools (e.g., additional lives saved or reductions in costs) is outside the scope of this article, but using these results can help understand whether the pricing of a given offering is reasonable. For example, if a vendor is offering a quote on continued pre-training, one would like to know how big the model is and how many passes they'll make on your institution's data to get an equivalent number of "runs" they are performing. Or, one could tell if a quote for a model trained from scratch offers a much larger model than an institution's data justifies.

*Deciding whether to train your own large language model*

Institutions with large data resources in cloud environments and human resources with AI expertise may be tempted to build their own LLMs using open-source infrastructure. If training from scratch, Table 2 can help decide whether the size of the institution's data resources justify a model large enough to bother. Even to train a 65 billion parameter model would require 4TB of text data, which is well beyond what many institutions will have, and the cost estimate is greater than $3 million per run. On the other hand, for less than $1 million, one could build on top of an existing 65 billion parameter model with more than 1TB of text data. Further, the benefit of standing up the infrastructure would allow continual training as data and evidence accumulates to allow model updating to reflect most recent patient practices and evidence-based medicine. We note that it is an open research question whether a model trained from scratch only on EHR data would be useful, or whether the massive amount of conventional linguistic data from the general domain is specifically important. Recent work has shown promising results for smaller models trained only on clinical text, though primarily for tasks requiring general medical knowledge [31].

*Whether to move now or to wait*

The excitement caused by these models might be considered a "hype wave" that could lead to eager institutions making investments on an expedited schedule. The results of our analyses could help temper enthusiasm and better ground decision-making about when to act. Technology is continuing to improve – see recent work that shows that models can be compressed significantly [32] – and the relative merits of training from scratch versus continued pre-training versus supervised fine tuning are still being adjudicated. Given this uncertainty, that compute costs will continue decreasing, and parameter-efficient tuning is an emerging technology, institutions with lesser data and financial resources may find the costs are too steep and the benefits too uncertain at this time. One compromise option for such an institution is to build a model at the smaller end of scale (e.g., 1.5 billion parameters) to build institutional knowledge around LLM training to be better prepared to build or evaluate larger models in the future.

*Important future directions*

While there are many groups already pursuing this kind of training, there are a number of unanswered research questions that are important for decision-making. 1) How important is data cleaning? There is a literature suggesting cleaning pre-training data is very valuable [33], [34] and EHR data is notoriously duplicative and unclean [35]. On the one hand, scale requirements may be lessened with higher quality data. On the other hand, removing junk tokens from EHR text will significantly reduce the amount of usable training data, perhaps increasing nominal data size requirements; 2) How many passes do we want the training process to take over the data? Most large models only see most tokens once, relying on the massive scale of pre-training data and repetition within the data of important information. For EHR LLM purposes, multiple passes may be required to fully integrate the relatively smaller data amounts into the model. Current work is exploring the question of how repeating data affects pre-training [36], [37] but not in an EHR setting. 3) What is the right data ratio? It is likely that some general domain text will be needed during pre-training, if only to give the model examples of what summarization, question-asking, and other linguistic phenomena look like. But exactly how much general domain data is required? 4) How frequently will these models need to be re-trained, and should model capacity consider the size of future data resources? and 5) What are the costs of human feedback and who will pay them? Instruction tuning [20] was used to train ChatGPT and is responsible for its "alignment" with user needs. While the cost of this part of the training pales in comparison to the pre-training costs, someone must still do the work to obtain high quality instruction tuned data sets. Recent work has repurposed medical language datasets to approximate an instruction tuning dataset [22] but it differs from the way real users might use the system. Obtaining high quality data will require expert clinicians using the system and giving feedback. This may be an opportunity for institutions to collaborate, as this scale of data can be more easily de-identified for sharing.

*Other costs*

The cost estimates in our results correspond to a single training iteration through a given amount of data. However, a full cost estimate requires accounting for human resources, data pre-processing time, model development iteration time, regular model updates, and many other factors. Many of these costs are difficult to estimate, but we can get a starting point to estimate model development iteration time from the LLaMA technical report. That paper reports a total compute time multiplier of 5.14 (e.g., total compute time is 5.14 times the cost of the final run required to train the 65 billion parameter model). Given the expertise at Meta AI, we consider that multiplier a lower bound.

*Cloud versus on-premises computing*

A limitation of the present work is that we do not quantify the trade-offs between cloud and on-premises GPU computing. Cloud computing offers flexibility, with the ability to scale to many parallel GPU instances, while on premises GPUs allow institutions to potentially amortize the upfront hardware costs across multiple LLM training runs, or on other research uses. However, our view is that without the scalability of the cloud, training timelines are unreasonably long – LLM pre-training with even 8 A100s would take months to perform one training pass of even modest sized (13B parameter) LLMs.

## 6. Conclusion

This work has outlined important considerations related to data scale and computing costs that healthcare institutions must consider when making decisions about how to best acquire and deploy LLMs in their systems. Given the excitement about LLMs, understanding these issues and the resulting trade-offs can potentially ground discussions around their implementation in reality.

## Contributions

TM and WZ conceived the study and designed the experiments, with guidance from MA and DB. TM and WZ wrote the first draft. All authors contributed to the editing and approved the final version of the manuscript.

## Funding sources

TM is supported by NIH grants R01LM012973, R01MH126977, R01HL151604, and R01LM013486.

WZ is supported by NIH grant R01LM012973.

MA was supported by NIH grant R01DA051464.

## Data Availability

Most of the data used to carry out this work was publicly available as part of previous works (e.g., details of training of Llama models). Upon reasonable request, the spreadsheet we used to compile all of these variables into calculated tables will be shared. For modeling the relationship between dataset size and token counts, we used two internal datasets containing protected health information, and which we cannot share for patient privacy reasons. We also used MIMIC-III, which we accessed according to the terms of a data use agreement, which do not allow us to redistribute. However, this dataset can be accessed at the following url: https://physionet.org/content/mimiciii/1.4/.

## Competing Interests

The authors declare no Competing Financial or Non-Financial Interests.